\documentclass[runningheads]{llncs}
\usepackage{graphicx}
\usepackage{authblk}
\usepackage{graphicx}
\usepackage{subfigure}
\usepackage{array}
\usepackage{float}
\usepackage{booktabs}
\usepackage{bbding}
\usepackage{etoolbox}
\usepackage{color}
\usepackage{soul}
\usepackage{algorithm}
\usepackage{algorithmic}
\usepackage{cite}
\usepackage{balance}
\usepackage{amsmath, amssymb}
\usepackage{makecell}
\usepackage{multirow}
\usepackage{hyperref}

\begin{document}
\title{LighTDiff: Surgical Endoscopic Image Low-Light Enhancement with T-Diffusion}
\titlerunning{LighTDiff: Endoscopic Image Low-Light Enhancement}
%

%
\authorrunning{T. Chen et al.}

\author{Tong Chen\inst{1~\star} 
\and Qingcheng Lyu \inst{1~\star}
\and Long Bai\inst{2,3}
\thanks{Co-first authors.}
\and Erjian Guo\inst{1}
\and Huxin Gao\inst{2}
\and Xiaoxiao Yang\inst{4}
\and Hongliang Ren\inst{2,3}
\and Luping Zhou\inst{1} 
\thanks{Corresponding author.}
}
\institute{Faculty of Electrical Engineering, The University of Sydney, Sydney, Australia
\and Department of Electronic Engineering, The Chinese University of Hong Kong (CUHK), Hong Kong SAR, China
\and Shenzhen Research Institute, CUHK, Shenzhen, China
\and Qilu Hospital of Shandong University, Jinan, China\\
\email{tche2095@uni.sydney.edu.au, luping.zhou@sydney.edu.au}}
\maketitle              
\begin{abstract}
Advances in endoscopy use in surgeries face challenges like inadequate lighting. Deep learning, notably the Denoising Diffusion Probabilistic Model (DDPM), holds promise for low-light image enhancement in the medical field. However, DDPMs are computationally demanding and slow, limiting their practical medical applications. To bridge this gap, we propose a lightweight DDPM, dubbed LighTDiff. It adopts a T-shape model architecture to capture global structural information using low-resolution images and gradually recover the details in subsequent denoising steps. We further prone the model to significantly reduce the model size while retaining performance. While discarding certain downsampling operations to save parameters leads to instability and low efficiency in convergence during the training, we introduce a Temporal Light Unit (TLU), a plug-and-play module, for more stable training and better performance. TLU associates time steps with denoised image features, establishing temporal dependencies of the denoising steps and improving denoising outcomes. Moreover, while recovering images using the diffusion model, potential spectral shifts were noted. We further introduce a Chroma Balancer (CB) to mitigate this issue.  Our LighTDiff outperforms many competitive LLIE methods with exceptional computational efficiency.
Our code is available at
\href{https://github.com/DavisMeee/LighTDiff}{github.com/DavisMeee/LighTDiff}.
\end{abstract}
\section{Introduction}
\begin{figure}[t]
    \centering
    \includegraphics[width=1\linewidth, trim=0 200 0 0]{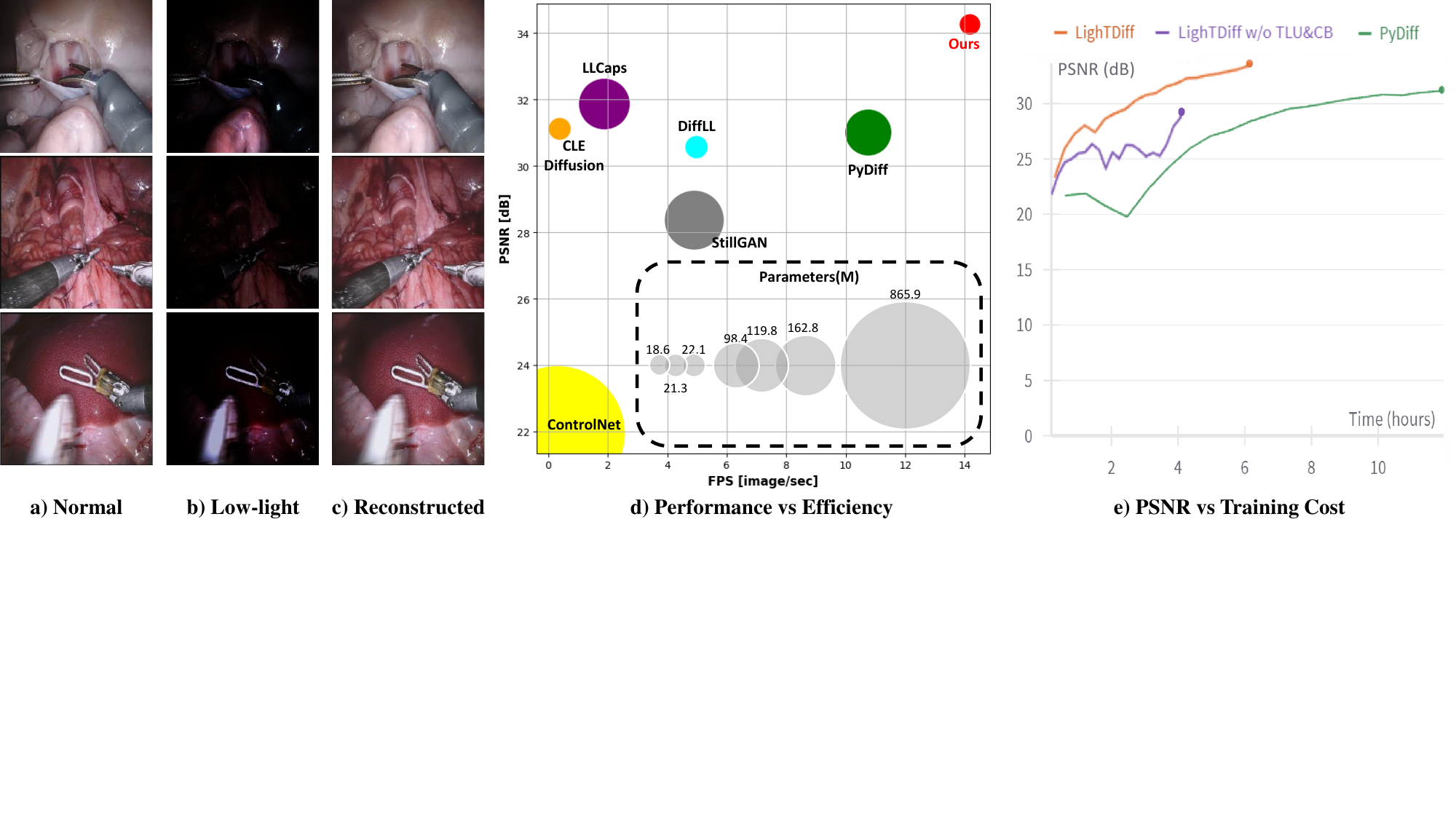}
    \caption{Image comparison for normal lighting (a), low lighting (b), and LighTDiff reconstruction (c); Model comparison in performance and efficiency (d), and comparison of training costs (i.e., training hours for the same number of total iterations) (e). 
    }
    \label{fig:illustration}
\end{figure}

Minimally invasive surgery (MIS), now a standard for various procedures~\cite{darzi2002recent}, offers advantages such as reduced trauma, faster recovery, and shorter hospital stays~\cite{rueckert2024methods}. Endoscopes, equipped with high-definition cameras and flexible maneuvering, provide surgeons with a clear field of vision for precise procedures~\cite{wang2023rethinking, bai2022transformer}. However, low-light conditions, as seen in (Fig.~\ref{fig:illustration} (b)), can complicate surgery due to poor illumination and contrast. Insufficient brightness affects detail recognition, making it challenging for surgeons to observe tissue structures or pathological areas. This complicates lesion localization, hampers fine manipulation, prevents effective guidance, and increases the risk of errors. An effective and efficient low-light image enhancement (LLIE) framework is essential to aid physicians in endoscopy-assisted MIS.

Conventional methods using hand-crafted image features~\cite{guo2016lime, liu2016contrast, li2018structure} for low-light imaging are often constrained by stringent assumptions, hampering their applicability. In contrast, recent advancements in deep learning~\cite{wang2022low, zamir2022MIRNetv2} allow models to automatically learn complex image features, enabling them to handle diverse lighting conditions and offer accurate enhancements. Advanced LLIE methods for medical endoscopy have emerged, such as  Gomez~\textit{et al.}\cite{gomez2019low} using a convolutional neural network trained on synthetic data for laryngoscope LLIE, Ma~\textit{et al.}\cite{ma2021structure} proposing a Generative Adversarial Network (GAN) model for low-light enhancement via unpaired training, and Bai~\textit{et al.}~\cite{2023llcaps} presenting a method using reverse diffusion for endoscopy image enhancement. Despite these efforts, efficiency and performance limitations persist, restricting their practical application in real-world endoscopy scenarios.

Recently, denoising diffusion probabilistic model (DDPM)~\cite{ho2020DDPM} has demonstrated strong capabilities in various generation tasks~\cite{podell2023sdxl, 2023controlnet}. Although existing works using DDPM for LLIE tasks~\cite{zhou2023pyramid,yin2023cle,jiang2023low} have shown promising performance, applying DDPM to enhance endoscopic images in low-light environments faces challenges. Firstly, existing diffusion models incur high computational costs, requiring significant computational resources and time, which is impractical for real-time applications such as endoscopy. Secondly, the endoscopic devices themselves have limited hardware capacity to deploy complex models.

To overcome these limitations, we propose \textbf{LighTDiff}, a lightweight diffusion architecture utilizing less than 50\% of the parameters of the original DDPM~\cite{ho2020DDPM}, making it suitable for consumer-grade hardware without compromising diffusion model performance. 
Our contributions are summarized as follows: \underline{First}, we adopt an inconstant resolution diffusion structure~\cite{zhou2023pyramid}, enabling efficient DDPM with iterative processes at varying resolution levels. Notably, we further optimize our model by strategically pruning certain downsampling components, resulting in a substantial reduction in size while maintaining performance. \underline{Second}, to address potential instability issues introduced by model pruning and improve performance, we introduce the Temporal Light Unit (TLU), a plug-and-play module, and employ it as the building block of the U-net backbone. TLU, conducting lay normalization, injects the time step into the image features of the current denoise stage. This temporal injection facilitates better contextual understanding, allowing the model to discern temporal dependencies of the denoising steps for improvement. \underline{Third}, we introduce the Chroma Balancer (CB), a corrective measure designed to adjust image channel bias distribution at each diffusion step, ensuring accurate diffusion outcomes. \underline{Fourth}, our extensive experiments conducted on two public datasets and a real-world dataset demonstrate promising performance and exceptional efficiency of our LighTDiff (Fig.~\ref{fig:illustration} (d)).

\section{Methodology}
\subsection{Preliminaries }

\textbf{Denoising Diffusion Probabilistic Model }(DDPM)~\cite{ho2020DDPM,saharia2022image} can be conceptualized as a forward adding noise and reverse denoising procedure.

This design enables the transformation of data from an observed space to a latent space characterized by Gaussian noise, followed by reconstruction to the original space, thus facilitating robust generative modeling. The diffusion model starts with a given data distribution $y_0$ $\in$ $q(y_0)$ then gradually adds Gaussian Noise ($\mathcal{\zeta}$) to it. Set the max time step as $T$. The step sizes are controlled by a variance schedule $\{\beta_{t}\in(0,1)\}_{t=1}^T$. We define $\mathcal{\gamma}_t = 1- \mathcal{\beta}_t$ and $\hat{\mathcal{\gamma}}_t = \prod \limits_{i=1}^t \mathcal{\gamma}_i$. The whole stage of the diffusion model can be formulated as:
\begin{equation}\label{equ:0}
\left\{
\begin{aligned}
  & q(y_t|y_{t-1})= \mathcal{N}(y_t,\sqrt{\gamma_t}y_{t-1},\beta_t\mathcal{\zeta})\\
  & p(y_{t-1}|y_{t})= \mathcal{N}(y_{t-1};\frac{\sqrt{\hat{\mathcal{\gamma}}_{t-1}}\mathcal{\beta}_t}{1-\mathcal{\hat{\gamma}}_t}f_{\theta}(y_t)+\frac{\sqrt{\gamma_t}(1-\hat{\gamma}_{t-1})}{1-\hat{\gamma}_t}y_t, \frac{1-\hat{\gamma}_{t-1}}{1-\hat{\gamma}_t}\beta_t\zeta)
\end{aligned}
\right.
\end{equation}
where $f_\theta(y_t)$ is an output $\tilde{y}_0$ that is estimated from denoising part $\mathcal{\zeta}_\theta(y_t)$. 
\subsection{Proposed Methodology}

\begin{figure*}[t]
    \centering
    \includegraphics[width=1\linewidth, trim=0 30 -10 0]{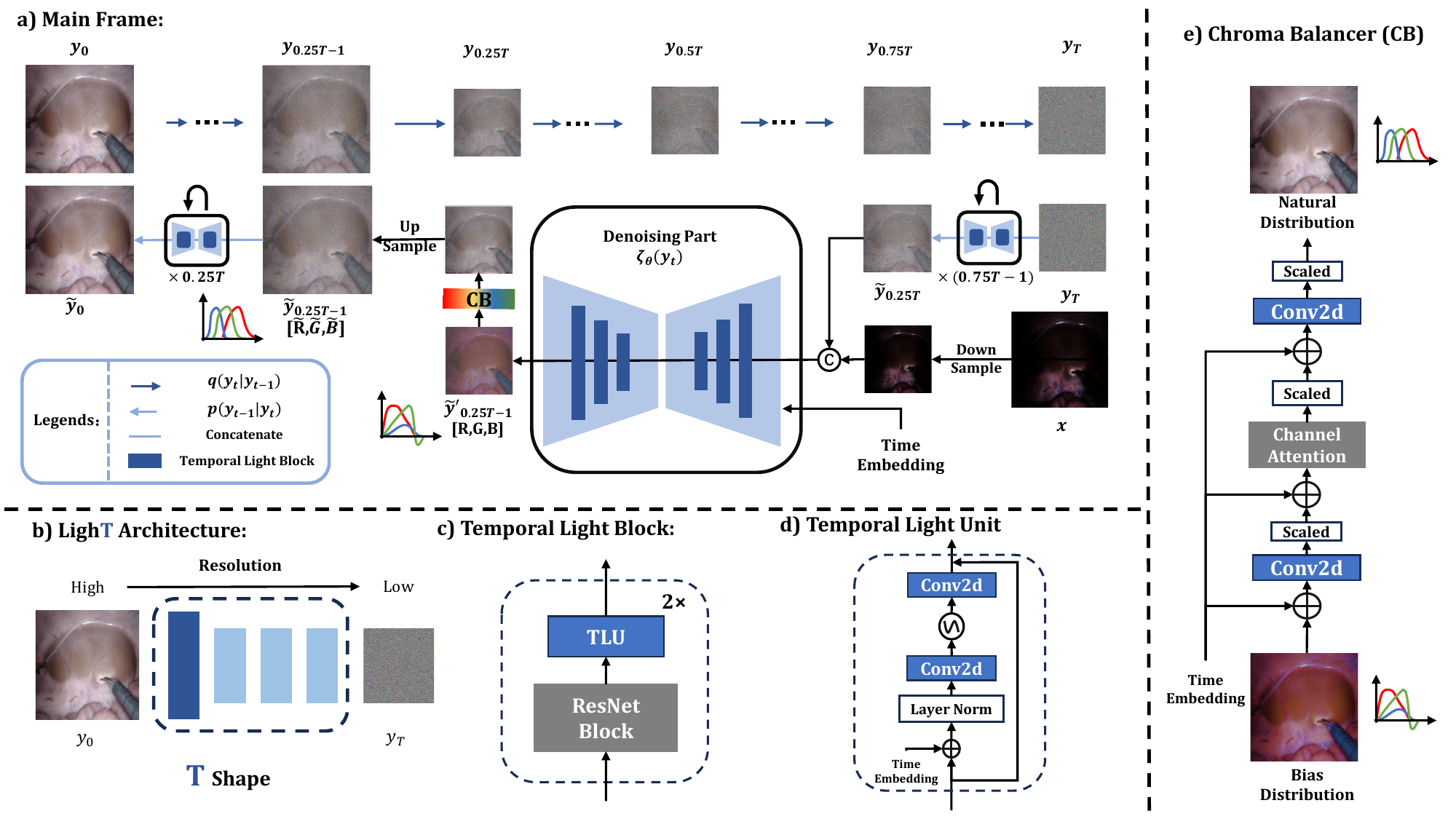}
    \caption{
    The overview of our proposed LighTDiff. Panel \textbf{a)} illustrates the entire process, where the original image $y_0$ undergoes noise diffusion to generate $y_t$, and the model learns to reconstruct the original image from different time steps. The denoised output ${\tilde{y}_0}$ is further adapted by the Chroma Balancer (CB) to approach a natural distribution, resulting in ${\tilde{y}'}_0$. Panel (b) is LighTDiff architecture. Panel (c) illustrates the Temporal Light Block, with the details of Temporal Light Unit (TLU) in Panel (d). The network structure of CB is given in Panel (e).
    }
    \label{fig:Schematric}
\end{figure*}

\subsubsection{LighT Architecture}
\label{sec:overall}

During the diffusion process, high-quality images are gradually transformed into low-quality images, which are then reconstructed in the reverse process. We find that when the noise is added to a certain extent, performing a resolution reduction and continuing to add noise at smaller scales will not significantly affect the quality of image recovery. Therefore, we designed our LighT architecture. The overview of our framework can be found in (Fig.~\ref{fig:Schematric}(a)). 
In this architecture, we reduce the resolution of the input after the quarter-journey. Due to the inconsistent resolution inputs, the general DDPM formula no longer fits in this case. The forward step has been adjusted as follows:
\begin{equation}
\begin{aligned}
  & q(y_t|y_{t-1})= \mathcal{N}(y_t;\sqrt{\mathcal{\gamma}_t}(y_{t-1}\downarrow_\frac{r_{t-1}}{r_t}),\beta_t\mathcal{\zeta}),\\
\end{aligned}
\end{equation}
and the Reverse Process can be expressed as:
\begin{equation}\label{equ:3}
p(y_{t-1}|y_{t})= \left\{ 
\begin{aligned}
 & \mathcal{N}(y_{t-1};\sqrt{\hat{\gamma}_{t-1}}(y_{t}\uparrow_\frac{r_{t}}{r_{t-1}}),(1-\hat{\mathcal{\gamma}_t})\mathcal{\zeta}),~when~r_t < r_{t-1};~else\\
  & \mathcal{N}(y_{t-1};\frac{\sqrt{\hat{\mathcal{\gamma}}_{t-1}}\mathcal{\beta}_t}{1-\mathcal{\hat{\gamma}}_t}f_{\theta}(y_t)+\frac{\sqrt{\gamma_t}(1-\hat{\gamma}_{t-1})}{1-\hat{\gamma}_t}y_t, \frac{1-\hat{\gamma}_{t-1}}{1-\hat{\gamma}_t}\beta_t\zeta). 
\end{aligned}
\right.
\end{equation}
Here $r_t$ represents the image resolution at time step $t$. $\uparrow_\frac{r_{t}}{r_{t-1}}$ and $\downarrow_\frac{r_{t}}{r_{t-1}}$  represent up and down sampling adjustments based on resolution changes. It is noted that excessive downsampling may cause a loss of high-frequency information when doing the upsampling recovery~\cite{liu2023devil}.  To address this, we ensure that the model is downsampled at most twice in the denoising process, leading to a different diffusion schedule from~\cite{zhou2023pyramid}, contributing also to an overall lightweight model. Although pruning will significantly help speed up, it also causes an unstable training procedure, as shown by the purple curve in (Fig~\ref{fig:illustration}(e)).

\subsubsection{Temporal Light Block}
Temporal Light Block (TLB) is the core component of our denoising model, comprising a ResBlcok and Temporal Light Unit (TLU), as illustrated in (Fig.~\ref{fig:Schematric}~(c)). It is contrastive to the Attention ResBlock (ResBlock with a separate Self-Attention Block) employed as the building blocks of the vanilla DDPM. This latter focuses on the features of the current step for denoising and demands significant computing resources. Compared with the self-attention block, TLU is a lightweight plug-and-play block. It injects the time step $t$ into the image features $\boldsymbol{\mathcal{F}}$ of the current denoise stage, enhancing contextual understanding and enabling the model to discern temporal dependencies of the denoising steps. The detail of TLU is shown in (Fig.~\ref{fig:Schematric}~(d)). The normalization layer ensures stability throughout training, followed by a `channel-level activation' to enhance feature extraction.

The TLU is formulated as:
\begin{equation}
    \begin{aligned}
    \mathbf{TLU}(\boldsymbol{\mathcal{F}},t) = \boldsymbol{{c^{-1}}}(\boldsymbol{\sigma}(\boldsymbol{c}(\boldsymbol{norm}(\boldsymbol{\mathcal{F}}+t)))+\boldsymbol{\mathcal{F}},
    \end{aligned}
    \label{equ:4}
\end{equation}
where $\boldsymbol{c}(\cdot)$ and $\boldsymbol{c}^{-1}(\cdot)$ represent convolution and deconvolution operations, respectively, while $\boldsymbol{\sigma}(\cdot)$ denotes non-linear activation. This design circumvents the computationally costly self-attention while maintaining good performance. Moreover, the time injection provides temporal context for better denoising.  

\subsubsection{Chroma Balancer}
In the later stages, random sampling and inaccurate denoising may introduce color bias and exposure shifts, leading to color distribution distortion~\cite{li2023alleviating, chen2022speed}. While Zhou\textit{et al.}\cite{zhou2023pyramid} employed a global corrector using time embedding and scaling adaption to address this issue, their performance is limited due to the neglect of channel connections. To overcome this limitation, we introduce a Chroma Balancer (CB) module to rectify the bias distribution toward a natural distribution, as illustrated on the right in (Fig.\ref{fig:Schematric} (e)). 

In contrast to the global corrector in~\cite{zhou2023pyramid}, our approach incorporates channel attention. This allows focused emphasis on key dimensions while suppressing irrelevant distractions. As a result, CB substantially improves performance with minimal additional computational overhead.

\subsubsection{Overall Structure}
An overview of our framework can be found in (Fig.~\ref{fig:Schematric} (e)). Our LighTDiff contains a denoising component constructed by Temporal Light Blocks (TLB) (including ResBlock and TLU) and a Chroma Balancer (CB).
We use the SmoothL1 loss~\cite{girshick2015fast} to optimize denoising and CB with no additional optimization objectives. The $\varepsilon$ is set to $1$ empirically.
\begin{equation}
    \mathcal{L}_{smoothL1}=\left\{
    \begin{aligned}
    &0.5(y_0-\tilde{y}_0)^2/\varepsilon      &\hfill \qquad if |y_0-\tilde{y}_0|<\varepsilon\\
    &|y_0-\tilde{y}_0|-0.5\varepsilon   & otherwise
    \end{aligned}
    \right.
    \label{equ:loss}
\end{equation}

\section{Experiments}
\label{sec:exper}

\subsection{Dataset}
\label{sec:dataset}

Given the challenge of obtaining real paired endoscopic low-light and normal-light images, we synthesize low-light images based on the normal light images in the EndoVis17~\cite{allan2019endovis17} and the EndoVis18~\cite{allan2020endovis18} datasets to obtain corresponding low-light and normal-light pairs and further applied our model to enhance images in a real-world low-light endoscopic submucosal dissection (ESD) surgery dataset.

\textbf{EndoVis17}~\cite{allan2019endovis17} and \textbf{EndoVis18}~\cite{allan2020endovis18} are two publicly accessible datasets for surgical instrument segmentation. EndoVis18 and EndoVis17 contain 2,400 and 2,235 images selected from 10 and 14 videos, respectively. EndoVis17 and EndoVis18 dataset is split into 1800 and 1639 training images and 1200 and 596 test images following\cite{gonzalez2020isinet}. We resized them into $256\times256$ resolution. To synthesize low-light images, we adopted random Gamma and illumination reduction following~\cite{lore2017llnet,li2021low,2023llcaps}.

\textbf{Real-world dataset} is an in-house LLIE dataset collected from 20 ESD surgery videos on pigs (Approval No. DWLL-2021-021)~\cite{gao2024transendoscopic}. We manually collected 61 low-light images with segmentation labels of the instruments and the backgrounds. All tags are hand-labeled and corrected by doctors. Due to the anonymous policy, the animal study ethical approval is temporarily concealed.

\subsection{Comparison Methods and Evaluation Metrics}
\label{sec:implementation}
We benchmark our approach against traditional methods like LIME~\cite{guo2016lime} and DUAL~\cite{zhang2019dual}, typical CNN and RCNN-based methods such as Zero-DCE~\cite{guo2020zero}, SNR-Aware~\cite{xu2022snr}, MIRNet~\cite{zamir2020MIRNetv1}, and MIRNetv2~\cite{zamir2022MIRNetv2}, and GAN-based ones like EnlightenGAN~\cite{jiang2021enlightengan} and StillGan~\cite{ma2021structure}. We also compare with DDPM-based models like LLCaps~\cite{2023llcaps}, DiffLL~\cite{jiang2023low}, CLEDiffusion~\cite{yin2023cle}, ControlNet~\cite{2023controlnet}, and PyDiff~\cite{zhou2023pyramid}.

For evaluation, we use common image quality assessment metrics, including Peak Signal-to-Noise Ratio (PSNR)\cite{huynh2008scope}, Structural Similarity Index (SSIM)\cite{wang2004image} and Learned Perceptual Image Patch Similarity (LPIPS)\cite{zhang2018unreasonable}. Additionally, we assess the impact of image enhancement on downstream segmentation tasks using both the synthetic EndoVis17 dataset and our real low-light image dataset. The segmentation model, utilizing a pre-trained ResNet-101\cite{he2016deep} as the backbone and DeeplabV2~\cite{chen2017deeplab} as the decoder, was trained on the EndoVis17 training set for 200 epochs. It was directly applied to the enhanced images in the real low-light dataset without fine-tuning. The segmentation results are evaluated using mean Intersection over Union (mIoU) and Dice similarity coefficient (Dice).
\subsection{Results}
\label{sec:results}

\begin{table}[t]
	\caption{
 Performance comparison by image qualities and downstream segmentation tasks. The image quality metrics are only reported for the synthetic data as the real-world data lack ground truth. The segmentation performance on the real-world dataset is obtained by directly applying a segmentation model trained on the EndoVis17 training set without any finetuning.
	}
 	\centering
	\label{tab:1}  
\resizebox{\textwidth}{!}{	
\begin{tabular}{c|c|ccc|ccc|cc|cc}
\hline

\multirow{2}{*}{Models}&\multicolumn{1}{c|}{Efficiency} & \multicolumn{3}{c|}{EndoVis17}          & \multicolumn{3}{c|}{EndoVis18} & \multicolumn{2}{c|}{EndoVis17 Seg} & \multicolumn{2}{c}{Real-world Seg}  \\ \cline{2-12} 
&FPS$\uparrow$ & PSNR $\uparrow$ & SSIM $\uparrow$  & LPIPS $\downarrow$ & PSNR $\uparrow$ & SSIM $\uparrow$  & LPIPS $\downarrow$ & Dice $\uparrow$ & mIoU$\uparrow$ & Dice $\uparrow$ & mIoU $\uparrow$  \\ \hline 
LIME~\cite{guo2016lime}                   &1.54 & 11.56 & 24.20 & 0.3973 & 11.69 & 29.91 & 0.3848 & 33.39 & 24.90 & 48.79 & 35.29 \\
DUAL~\cite{zhang2019dual}                 &1.31 & 11.50 & 27.29 & 0.4067 & 11.52 & 36.38 & 0.3966 & 39.70 & 27.39 & 54.62 & 40.45 \\
Zero-DCE~\cite{guo2020zero}               &14.96 & 11.23 & 31.34 & 0.4980 & 10.90 & 37.27 & 0.5128 & 21.32 & 13.29 & 22.92 & 14.17 \\
EnlightenGAN~\cite{jiang2021enlightengan} &\textbf{20.69} &24.97 & 83.55 & 0.1210 & 22.79 & 81.71 & 0.1604 & 71.71 & 58.84 & 49.14 & 35.62\\
MIRNet~\cite{zamir2020MIRNetv1}           &3.15 &31.69 & 93.03 & 0.0673 & 29.08 & 92.81 & 0.0785 & 78.84 & 66.66 & 57.99 & 42.98 \\
MIRNetv2~\cite{zamir2022MIRNetv2}         &5.38 &31.94 & 93.48 & 0.0671 & 28.96 & 92.63& 0.0799 & 82.87 & 72.27 & 58.12 & 43.64  \\
SNR-Aware~\cite{xu2022snr}        &12.32 &27.37 &90.57 & 0.1149 & 26.57 & 89.79 & 0.1335 & 75.72 & 63.33 & 58.87 & 44.35\\
StillGan~\cite{ma2021structure}         &4.90 & 28.39 & 90.28 & 0.0920 & 26.78 & 88.27 & 0.1221 & 82.57 & 71.61 & 59.14 & 44.30  \\
LLCaps~\cite{2023llcaps}              &1.87  & 31.88 & 93.53 & 0.0589 & 25.18 & 90.16 & 0.1110 & 83.49 & 73.07 & 57.34 & 42.90 \\
DiffLL~\cite{jiang2023low}             &4.98 & 30.57 & 92.99 & 0.0763 & 28.07 & 91.82 & 0.0962 & 83.39 & 72.99 & 56.61 & 41.73 \\
CLEDiff~\cite{yin2023cle} &0.37 & 31.12 & 94.54 & 0.0418 & 29.66 & 82.38 & 0.0865 & 82.44 & 71.49 & 56.62 & 41.91 \\
ControlNet~\cite{2023controlnet}         &0.30 & 21.97 & 64.82 & 0.2179 & 22.15 & 68.68 & 0.2565  & 75.20 & 62.62 & 55.34 & 40.23  \\
PyDiff~\cite{zhou2023pyramid}         &10.74  & 31.02 & 94.50 & 0.0510 & 29.37 & 93.82 & 0.0711  & 84.81 & 74.85 & 52.66 & 37.64  \\
\textbf{LighTDiff (Ours)} &14.19 & \textbf{34.28} & \textbf{95.72} & \textbf{0.0325} & \textbf{31.99} & \textbf{94.91} & \textbf{0.0531} & \textbf{86.65} & \textbf{75.97} & \textbf{59.55} & \textbf{44.57}  \\ \hline
\end{tabular}}
\end{table}

\begin{figure*}[!t]
    \centering
    \includegraphics[width=0.95\linewidth, trim=0 70 0 20]{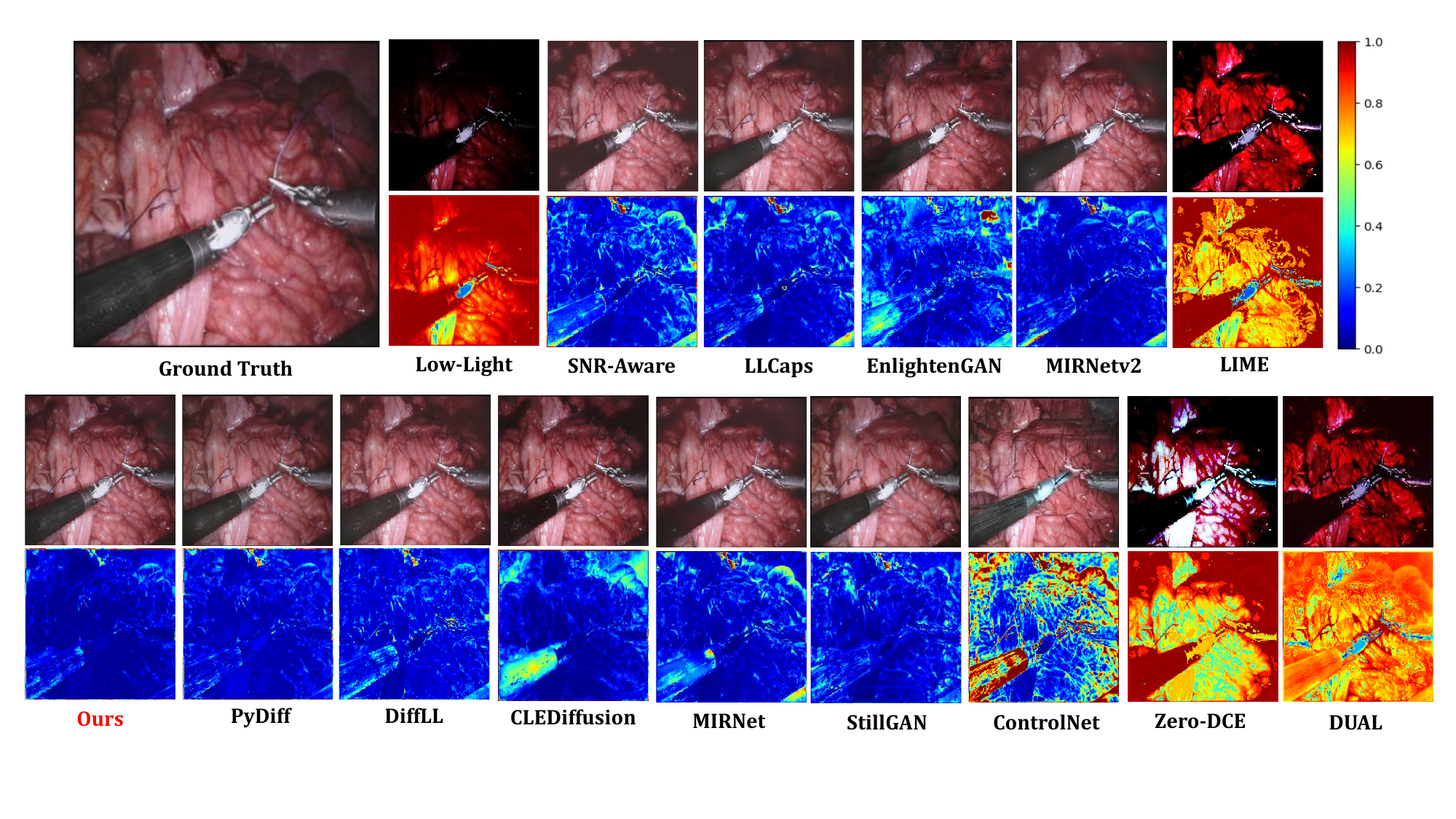}
    \caption{The quantitative results for LighTDiff compared with SOTA approaches on EndoVis17~\cite{allan2019endovis17}. The first row shows the enhanced images for different LLIE baselines, and the second row shows the reconstruction error heat maps.  Blue to red indicates the error from small to large. Zoom to see the details.
    }
    \label{fig:IMG_visual}
\end{figure*}

\noindent\textbf{Image Enhancement} Table~\ref{tab:1} quantitatively compares LighTDiff's performance with existing methods. Conventional approaches like LIME~\cite{guo2016lime} and DUAL~\cite{zhang2019dual} yield unsatisfactory results due to strict assumptions. ControlNet~\cite{2023controlnet}, reliant on extensive training data, struggles with limited training samples and achieves suboptimal image restoration. In comparison to other state-of-the-art diffusion-based methods, LighTDiff demonstrates superior performance. Specifically, it outperforms two closer competitors CLEDiffusion~\cite{yin2023cle} by 3.16 dB on EndoVis17 and 2.33 dB on EndoVis18, and PyDiff~\cite{zhou2023pyramid} by 3.26 dB on EndoVis17 and 2.62 dB on EndoVis18, respectively, in PSNR. The SSIM of LighTDiff improves to 95.72\% in EndoVis17 and 94.91\% in EndoVis18. Our method's enhanced temporal context through TLU and adoption of CB contribute to superior image restoration quality. Qualitative results, along with error heatmaps, are visualized in Fig.~\ref{fig:IMG_visual} on the EndoVis17 dataset.

\noindent\textbf{Downstream Tasks} 
Furthermore, our LighTDiff is evaluated by downstream segmentation tasks to investigate the potential utility of the restored images. As depicted in Table~\ref{tab:1}, LighTDiff outperforms competitors in instrument segmentation, underscoring its superior efficacy in segmenting instrumental objects. It consistently surpasses all SOTA approaches in comparison in terms of mIoU and Dice scores on both synthetic and real-world datasets, highlighting its superior capability in low-light image restoration and edge preservation. 

\begin{table}[t]
\centering
	\caption{
        Ablation experiments of our LighTDiff on the EndoVis17 Dataset~\cite{allan2019endovis17}. To observe the performance changes, we (i) remove the LighT Architecture, (ii) degenerate the Chroma Balancer, (iii) remove the Temporal Light Unit, and (iv) discard time embedding from TLU.
	}
 	\centering
        \resizebox{\textwidth}{!}{
 \label{tab:ablation}
\begin{tabular}{c|c|c|c||p{1.5cm}<{\centering} p{1.5cm}<{\centering} p{1.5cm}<{\centering} p{1.5cm}<{\centering} p{1.5cm}<{\centering}}
\hline
\multicolumn{1}{c|}{\makecell[c]{LighT\\Architecture }}  & \multicolumn{1}{c|}{\makecell[c]{Chroma\\ Balancer}} & \multicolumn{1}{c|}{\makecell[c]{Temporal\\Light Unit}}& \multicolumn{1}{c|}{\makecell[c]{Time\\ Embedding}} &Param(M)$\downarrow$  &FPS$\uparrow$& PSNR $\uparrow$ & SSIM $\uparrow$  & LPIPS $\downarrow$  \\ \hline
\XSolidBrush      &\XSolidBrush      & \XSolidBrush & \multicolumn{1}{c|}{-}    &98.43  &10.74  & 31.12 & 94.50 & 0.0510  \\
\Checkmark      &\XSolidBrush      & \XSolidBrush  & \multicolumn{1}{c|}{-}     &14.09 &15.18  & 29.341 & 94.03 & 0.0490  \\
\XSolidBrush       & \Checkmark     &\XSolidBrush & \multicolumn{1}{c|}{-}  &98.42  &10.81       & 33.39 & 95.35 & 0.0433 \\
 \XSolidBrush      &\XSolidBrush      &\Checkmark & \multicolumn{1}{c|}{\Checkmark}    &127.06  &7.27     & 32.69 & 95.34 & 0.0409 \\
\Checkmark       & \Checkmark     &\XSolidBrush & \multicolumn{1}{c|}{-}     &\textbf{14.08}   &\textbf{15.33}    & 31.97 &95.21 &0.0407 \\
\Checkmark      &\XSolidBrush   & \Checkmark    & \multicolumn{1}{c|}{\Checkmark}     & 18.59 &13.83     & 33.46 & 95.55 & 0.0361  \\
\XSolidBrush       & \Checkmark   & \Checkmark  & \multicolumn{1}{c|}{\Checkmark}  &127.05  &8.15  & 33.19 & 95.65 &0.0382 \\
\Checkmark      & \Checkmark   & \Checkmark  & \multicolumn{1}{c|}{\XSolidBrush}  &18.40  &14.18  & 32.54 & 95.58 &0.0345 \\
\Checkmark     & \Checkmark   & \Checkmark   &\multicolumn{1}{c|}{\Checkmark}   &18.58  &14.16   & \textbf{34.28} & \textbf{95.72} & \textbf{0.0325}  \\ \hline  
\end{tabular}}
\end{table}

\noindent\textbf{Computational Efficiency} 
To assess computational efficiency, we report the frames per second (FPS) in the first column of Table~\ref{tab:1}. Our method achieves significantly higher FPS compared to other high-performing methods, except for EnlightenGAN~\cite{jiang2021enlightengan}. However, EnlightenGAN's performance (PSNR 24.97/22.79 on EndoVis17/EndoVis18) is notably inferior to ours (34.28/31.99). Additionally, we present a visual comparison in (Fig~\ref{fig:illustration} (c)), illustrating the comprehensive efficiency and performance metrics of our LighTDiff in comparison to other diffusion-based models and StillGAN~\cite{ma2021structure}. Our model is positioned at the top-right corner with smaller diameters, indicating higher PSNR, FPS, and fewer parameters. Specifically, compared with PyDiff~\cite{zhou2023pyramid}, our model is 85.6\% lighter on parameters, 32\% faster on FPS, and the training cost (i.e., training time for the same number of total iterations) is nearly halved (Fig~\ref{fig:illustration} (d)).

\noindent\textbf{Ablation Study} 
Our ablation study on the EndoVis17 dataset highlights the strategic improvements introduced by our approach. Table~\ref{tab:ablation} evaluates both model parameters and performance, with the first row representing a baseline model like PyDiff. The LighT architecture significantly reduces model parameters to 14.08M parameters and increases speed to 15.27 FPS. However, this improvement is accompanied by a slight performance drop and an unstable training procedure as shown by the purple curve in (Fig~\ref{fig:illustration} (d)). Introducing TLU enhances (PSNR/SSIM) from (29.34/94.03) to (33.46/95.55) with a minor parameter increase. Our CB module maintains similar parameter sizes to the baseline, slightly increases speed, and significantly enhances PSNR/SSIM from (29.34/94.03) to (31.97/95.21). Integrating these three components, our complete LighTDiff achieves the best image quality while being significantly more efficient than the baseline. Additionally, removing the time embedding in our TLU results in a drop in PSNR/SSIM from (34.78/95.72) to (32.54/95.58), justifying the necessity of time injection.

\section{Conclusion}
\label{sec:conclusion}
In this study, we introduce LighTDiff, a lightweight endoscopy LLIE diffusion model. LighTDiff incorporates the Temporal Light Block (TLB) for improved denoising and training stability, the Chroma Balancer (CB) to address chroma bias, and the LighT Architecture ensuring swift inference without compromising restoration quality. Comparative studies demonstrate superior performance in image quality and speed, indicating its suitability for consumer-grade hardware. Future plans involve adapting the model for various medical applications, including real-time augmentation and surgical navigation.

%
%
%
%
\bibliography{reference}{}

\begin{thebibliography}{10}
\providecommand{\url}[1]{\texttt{#1}}
\providecommand{\urlprefix}{URL }
\providecommand{\doi}[1]{https://doi.org/#1}

\bibitem{allan2020endovis18}
Allan, M., Kondo, S., Bodenstedt, S., Leger, S., Kadkhodamohammadi, R., Luengo, I., Fuentes, F., Flouty, E., Mohammed, A., Pedersen, M., et~al.: 2018 robotic scene segmentation challenge. arXiv preprint arXiv:2001.11190  (2020)

\bibitem{allan2019endovis17}
Allan, M., Shvets, A., Kurmann, T., Zhang, Z., Duggal, R., Su, Y.H., Rieke, N., Laina, I., Kalavakonda, N., Bodenstedt, S., et~al.: 2017 robotic instrument segmentation challenge. arXiv preprint arXiv:1902.06426  (2019)

\bibitem{2023llcaps}
Bai, L., Chen, T., Wu, Y., Wang, A., Islam, M., Ren, H.: Llcaps: Learning to illuminate low-light capsule endoscopy with curved wavelet attention and reverse diffusion. In: International Conference on Medical Image Computing and Computer-Assisted Intervention. pp. 34--44. Springer (2023)

\bibitem{bai2022transformer}
Bai, L., Wang, L., Chen, T., Zhao, Y., Ren, H.: Transformer-based disease identification for small-scale imbalanced capsule endoscopy dataset. Electronics  \textbf{11}(17), ~2747 (2022)

\bibitem{chen2022speed}
Chen, G.: Speed up the inference of diffusion models via shortcut mcmc sampling. arXiv preprint arXiv:2301.01206  (2022)

\bibitem{chen2017deeplab}
Chen, L.C., Papandreou, G., Kokkinos, I., Murphy, K., Yuille, A.L.: Deeplab: Semantic image segmentation with deep convolutional nets, atrous convolution, and fully connected crfs. IEEE transactions on pattern analysis and machine intelligence  \textbf{40}(4),  834--848 (2017)

\bibitem{darzi2002recent}
Darzi, A., Mackay, S.: Recent advances in minimal access surgery. Bmj  \textbf{324}(7328),  31--34 (2002)

\bibitem{gao2024transendoscopic}
Gao, H., Yang, X., Xiao, X., Zhu, X., Zhang, T., Hou, C., Liu, H., Meng, M.Q.H., Sun, L., Zuo, X., et~al.: Transendoscopic flexible parallel continuum robotic mechanism for bimanual endoscopic submucosal dissection. The International Journal of Robotics Research  \textbf{43}(3),  281--304 (2024)

\bibitem{girshick2015fast}
Girshick, R.: Fast r-cnn. In: Proceedings of the IEEE International Conference on Computer Vision. p.~1440 (2015)

\bibitem{gomez2019low}
G{\'o}mez, P., Semmler, M., Sch{\"u}tzenberger, A., Bohr, C., D{\"o}llinger, M.: Low-light image enhancement of high-speed endoscopic videos using a convolutional neural network. Medical \& biological engineering \& computing  \textbf{57},  1451--1463 (2019)

\bibitem{gonzalez2020isinet}
Gonz{\'a}lez, C., Bravo-S{\'a}nchez, L., Arbelaez, P.: Isinet: an instance-based approach for surgical instrument segmentation. In: International Conference on Medical Image Computing and Computer-Assisted Intervention. pp. 595--605. Springer (2020)

\bibitem{guo2020zero}
Guo, C., Li, C., Guo, J., Loy, C.C., Hou, J., Kwong, S., Cong, R.: Zero-reference deep curve estimation for low-light image enhancement. In: Proceedings of the IEEE Conference on Computer Vision and Pattern Recognition. p.~1780 (2020)

\bibitem{guo2016lime}
Guo, X., Li, Y., Ling, H.: Lime: Low-light image enhancement via illumination map estimation. IEEE transactions on image processing  \textbf{26}(2),  982--993 (2016)

\bibitem{he2016deep}
He, K., Zhang, X., Ren, S., Sun, J.: Deep residual learning for image recognition. In: Proceedings of the IEEE conference on computer vision and pattern recognition. p.~770 (2016)

\bibitem{ho2020DDPM}
Ho, J., Jain, A., Abbeel, P.: Denoising diffusion probabilistic models. Advances in Neural Information Processing Systems  \textbf{33},  6840--6851 (2020)

\bibitem{huynh2008scope}
Huynh-Thu, Q., Ghanbari, M.: Scope of validity of psnr in image/video quality assessment. Electronics letters  \textbf{44}(13),  800--801 (2008)

\bibitem{jiang2023low}
Jiang, H., Luo, A., Fan, H., Han, S., Liu, S.: Low-light image enhancement with wavelet-based diffusion models. ACM Transactions on Graphics  \textbf{42}(6),  1--14 (2023)

\bibitem{jiang2021enlightengan}
Jiang, Y., Gong, X., Liu, D., Cheng, Y., Fang, C., Shen, X., Yang, J., Zhou, P., Wang, Z.: Enlightengan: Deep light enhancement without paired supervision. IEEE transactions on image processing  \textbf{30},  2340--2349 (2021)

\bibitem{li2021low}
Li, C., Guo, C., Han, L., Jiang, J., Cheng, M.M., Gu, J., Loy, C.C.: Low-light image and video enhancement using deep learning: A survey. IEEE transactions on pattern analysis and machine intelligence  \textbf{44}(12),  9396--9416 (2021)

\bibitem{li2018structure}
Li, M., Liu, J., Yang, W., Sun, X., Guo, Z.: Structure-revealing low-light image enhancement via robust retinex model. IEEE Transactions on Image Processing  \textbf{27}(6),  2828--2841 (2018)

\bibitem{li2023alleviating}
Li, M., Qu, T., Sun, W., Moens, M.F.: Alleviating exposure bias in diffusion models through sampling with shifted time steps. arXiv preprint arXiv:2305.15583  (2023)

\bibitem{liu2023devil}
Liu, Y., Li, J., Pang, Y., Nie, D., Yap, P.T.: The devil is in the upsampling: Architectural decisions made simpler for denoising with deep image prior. In: Proceedings of the IEEE International Conference on Computer Vision. p. 12408 (2023)

\bibitem{liu2016contrast}
Liu, Y.F., Guo, J.M., Yu, J.C.: Contrast enhancement using stratified parametric-oriented histogram equalization. IEEE Transactions on Circuits and Systems for Video Technology  \textbf{27}(6),  1171--1181 (2016)

\bibitem{lore2017llnet}
Lore, K.G., Akintayo, A., Sarkar, S.: Llnet: A deep autoencoder approach to natural low-light image enhancement. Pattern Recognition  \textbf{61},  650--662 (2017)

\bibitem{ma2021structure}
Ma, Y., Liu, J., Liu, Y., Fu, H., Hu, Y., Cheng, J., Qi, H., Wu, Y., Zhang, J., Zhao, Y.: Structure and illumination constrained gan for medical image enhancement. IEEE Transactions on Medical Imaging  \textbf{40}(12),  3955--3967 (2021)

\bibitem{podell2023sdxl}
Podell, D., English, Z., Lacey, K., Blattmann, A., Dockhorn, T., M{\"u}ller, J., Penna, J., Rombach, R.: Sdxl: Improving latent diffusion models for high-resolution image synthesis. arXiv preprint arXiv:2307.01952  (2023)

\bibitem{rueckert2024methods}
Rueckert, T., Rueckert, D., Palm, C.: Methods and datasets for segmentation of minimally invasive surgical instruments in endoscopic images and videos: A review of the state of the art. Computers in Biology and Medicine p. 107929 (2024)

\bibitem{saharia2022image}
Saharia, C., Ho, J., Chan, W., Salimans, T., Fleet, D.J., Norouzi, M.: Image super-resolution via iterative refinement. IEEE Transactions on Pattern Analysis and Machine Intelligence  \textbf{45}(4),  4713--4726 (2022)

\bibitem{wang2023rethinking}
Wang, G., Bai, L., Wu, Y., Chen, T., Ren, H.: Rethinking exemplars for continual semantic segmentation in endoscopy scenes: Entropy-based mini-batch pseudo-replay. Computers in Biology and Medicine  \textbf{165},  107412 (2023)

\bibitem{wang2022low}
Wang, Y., Wan, R., Yang, W., Li, H., Chau, L.P., Kot, A.: Low-light image enhancement with normalizing flow. In: Proceedings of the AAAI conference on artificial intelligence. vol.~36, pp. 2604--2612 (2022)

\bibitem{wang2004image}
Wang, Z., Bovik, A.C., Sheikh, H.R., Simoncelli, E.P.: Image quality assessment: from error visibility to structural similarity. IEEE transactions on image processing  \textbf{13}(4),  600--612 (2004)

\bibitem{xu2022snr}
Xu, X., Wang, R., Fu, C.W., Jia, J.: Snr-aware low-light image enhancement. In: Proceedings of the IEEE/CVF Conference on Computer Vision and Pattern Recognition. p. 17714 (2022)

\bibitem{yin2023cle}
Yin, Y., Xu, D., Tan, C., Liu, P., Zhao, Y., Wei, Y.: Cle diffusion: Controllable light enhancement diffusion model. In: Proceedings of the 31st ACM International Conference on Multimedia. pp. 8145--8156 (2023)

\bibitem{zamir2020MIRNetv1}
Zamir, S.W., Arora, A., Khan, S., Hayat, M., Khan, F.S., Yang, M.H., Shao, L.: Learning enriched features for real image restoration and enhancement. In: Computer Vision--ECCV 2020: 16th European Conference, Proceedings, Part XXV 16. pp. 492--511. Springer (2020)

\bibitem{zamir2022MIRNetv2}
Zamir, S.W., Arora, A., Khan, S., Hayat, M., Khan, F.S., Yang, M.H., Shao, L.: Learning enriched features for fast image restoration and enhancement. IEEE transactions on pattern analysis and machine intelligence  \textbf{45}(2),  1934--1948 (2022)

\bibitem{2023controlnet}
Zhang, L., Rao, A., Agrawala, M.: Adding conditional control to text-to-image diffusion models. In: Proceedings of the IEEE/CVF International Conference on Computer Vision. p.~3836 (2023)

\bibitem{zhang2019dual}
Zhang, Q., Nie, Y., Zheng, W.S.: Dual illumination estimation for robust exposure correction. In: Computer Graphics Forum. vol.~38, pp. 243--252 (2019)

\bibitem{zhang2018unreasonable}
Zhang, R., Isola, P., Efros, A.A., Shechtman, E., Wang, O.: The unreasonable effectiveness of deep features as a perceptual metric. In: Proceedings of the IEEE Conference on Computer Vision and Pattern Recognition. p.~586 (2018)

\bibitem{zhou2023pyramid}
Zhou, D., Yang, Z., Yang, Y.: Pyramid diffusion models for low-light image enhancement. arXiv preprint arXiv:2305.10028  (2023)

\end{thebibliography}
\bibliographystyle{splncs04}

\end{document}